\documentstyle[art10]{article}

\newtheorem{theorem}{Theorem}
\def\proof{\noindent{\bf Proof}}
\def\example{\noindent{\bf Example\ }}
\def\df{\noindent{\bf Definition}}

\def\[{\begin{equation}}
\def\]{\end{equation}}

\def\one{{\bf1}}

\def\dsl{\raise.15ex\hbox{/}\kern-.57em\partial}  

\def\ss#1#2{\mathrel{\mathop{#2}\limits^{#1}}}

\def\dal{\Box}

\def\4{$4\!\times\!4$}
\def\Ç{$\gamma$}

\def\M{{\cal M}}

\def\R{{\cal R}}
\def\C{{\cal C}}

\def\G{{\cal G}}
\def\L{{\cal L}}

\def\com{{\rm com}}

\def\diag{{\rm diag}}
\def\tr{{\rm tr}}
\def\U{{\rm U}}

\def\SU{{\rm SU}}

\def\com{{\rm com}}
\def\u{{\rm u}}

\def\mlc{\M(l,\C)}
\def\rr#1{(\ref{#1})}

\title{Gauge fields of the matrix Dirac equation
        \thanks{This work is supported by the Russian Fund of
        Fundamental Research, grant 95-01-00433Á}}

\author{N.G.Marchuk \thanks{
Steklov Mathematical Institute;
Moscow 117966,
Gubkina st. 8;
nikolai@marchuk.mian.su}}

\begin{document}

\maketitle

\begin{abstract}
We introduce an equation named matrix Dirac equation
which can be considered as a generalization of Dirac
equation for an electron.
The  liaison between  matrix Dirac equation
and standard Dirac equation is discussed.
We write a lagrangian from which
matrix Dirac equation can be derived. This lagrangian is
invariant under global unitary transformations of variables.
The requirement of a local (gauge) invariance of lagrangian leads
us to lagrangian with gauge fields.
\end{abstract}

\section*{Introduction.}
After the famous Dirac equation
for an electron was found in 1928 \cite{dirac}, works of H.Weil
\cite{weil},
V.A.Fock \cite{fock} and others appeared,  in which electromagnetic  field
was  described  as  gauge field of the Dirac equation appearing from
the demand of local gauge invariance with respect to phase transformation
(an abelian $\U(1)$ gauge group ) of Dirac`s lagrangian.  Further development
of this approach for the non-Abelian gauge fields was made  in
the  work  of  Yang  and Mills (1954) \cite{yang} considering the group of
isotopic transformations (gauge group $\SU(2)$).  Their work was  soon
generalized for the wide class of Lie groups. Non-Abelian gauge fields
began to be named Yang-Mills fields, whereas the equations describing
them  were named Yang-Mills equations.  In modern physics such fields
are used in models of electroweak and strong interactions.

     In the present article an equation named matrix Dirac equation
is introduced.
This equation can be considered as a generalization of Dirac
equation for an electron.  Certain features of matrix Dirac equation are
investigated (currents, canonical forms). Further on the basis
of local gauge invariance regarding unitary group a system of  equations
is  introduced consisting of matrix Dirac equation and equations
of Yang-Mills or Maxwell. This system of equations describes Dirac`s
field  interacting  with
the gauge  field  of  Yang-Mills or Maxwell.

     Some ideas  proposed  in  \cite{marchuk1,marchuk2,marchuk3}
are used in the article.  Certain
elements of the construction proposed could be found in the  works  of
Hestenes \cite{hestenes}, K\"ahler \cite{kahler}, Pestov
\cite{pestov}, Pezzaglia and Differ \cite{pezzaglia}.

\section{A standard Dirac equation.}
Let us consider a Dirac equation and its generalizations.
A vector $x=(x^0,x^1,x^2,x^3)\in \R^4$
defines a point in space-time, $x^0$ -- time coordinate,
$x^1,x^2,x^3$ -- space
coordinates, $\partial_\mu={\partial/{\partial x^\mu}},\ \mu=0,1,2,3$
are the partial derivatives and
$\dal=\partial_\mu \partial^\mu=
{\partial_0}^2-{\partial_1}^2-{\partial_2}^2-{\partial_3}^2$.
 A Klein-Gordon equation for
a function $\phi=\phi(x)$
\[
(\dal + m^2)\phi=0, \label{1.1}
\]
where $m$ -- nonnegative real number, describes particles with spin
$0$ and mass $m$. For a description of spin $1/2$ particles (electrons)
P.~A.~M.~Dirac suggested a system of equations of first order
that he received by factorizing the Klein-Gordon operator
\[
(i\gamma^\mu \partial_\mu + m)(i\gamma^\nu \partial_\nu - m)=-(\dal + m^2),
\label{1.2}
\]
where $\gamma^\mu,\ \mu=0,1,2,3$
are algebraic objects that satisfy the relations
\[
\gamma^\mu \gamma^\nu + \gamma^\nu \gamma^\mu = 2 g^{\mu\nu},
\quad \mu,\nu=0,1,2,3 \label{1.3}
\]
with Minkowski tensor $g=(g^{\mu\nu})={\rm diag}(1,-1,-1,-1)$.
$\gamma^\mu$ can be represented by the matrices of order no
less than four. We shall use the following representation for $\gamma^\mu$:
\[
\gamma^0=\pmatrix{\sigma^0 & 0 \cr 0 & -\sigma^0}, \quad
\gamma^k=\pmatrix{0 & -\sigma^k \cr \sigma^k & 0}, \quad k=1,2,3
\label{1.4}
\]

$$
\sigma^0=\pmatrix{1&0\cr 0&1},\quad
\sigma^1=\pmatrix{0&1\cr 1&0},\quad
\sigma^2=\pmatrix{0&-i\cr i&0},\quad
\sigma^3=\pmatrix{1&0\cr 0&-1},
$$
which is called the Dirac representation ($\sigma^k$ -- Pauli matrices).
Let $\one$ denote the identity \4-matrix and
$\gamma^{\mu\nu}=\gamma^\mu \gamma^\nu$ for $0\leq\mu<\nu\leq3$,
$\gamma^{\mu\nu\lambda}=\gamma^\mu \gamma^\nu \gamma^\lambda$ for
$0\leq\mu<\nu<\lambda\leq3$,
$\gamma^5=\gamma^{0123}=\gamma^0 \gamma^1 \gamma^2 \gamma^3$.
The 16 matrices
\begin{equation}
\one, \ \gamma^\mu, \ \gamma^{\mu\nu}, \ \gamma^{\mu\nu\lambda}, \ \gamma^5
\label{1.5}
\end{equation}
are linear independent and form a basis of ${\M}(4,{\C})$ -- algebra of four
dimensional complex matrices.

The Dirac equation has a form
\[
(i\gamma^\mu \partial_\mu-m \one)\psi=0, \label{1.6}
\]
and if we consider the aggregate in brackets as a \4-matrix, then $\psi=\psi(x)$
 must be
a matrix with four lines and an arbitrary number of columns. The identity
(\ref{1.2}) asserts that all
the components of the matrix $\psi$ (if they are smooth
enough) satisfy the Klein-Gordon equation.

We shall call the equation (\ref{1.6}) with one column matrix $\psi$
the standard Dirac equation. In that case
$\psi$ is called bispinor or Dirac spinor.
  One can also consider the
equations (\ref{1.6}) with $l>1$ columns in the matrix $\psi$. For the different
purposes physicists use the Dirac equation (\ref{1.6})
with the different numbers
of columns in $\psi$. The following list does not lay claim for completeness or
indisputability:
\begin{description}
\item[$l=1$:] Quantum electrodynamics (a gauge field theory with $\U(1)$
symmetry group).
\item[$l=2$:] Theory of electroweak interactions ($\SU(2)$ gauge field
theory).
\item[$l=3$:] Theory of strong interactions -- quantum chromodynamics
( $\SU(3)$ gauge field theory).
\item[$l=4$:] The Dirac equation (\ref{1.6}) with $l=4$ is
called a Rarita-Schwinger
equation. It is used for a description of spin $3/2$ particles.
\item[$l=5$:] Georgi and Glashow \cite{georgi}
 have suggested $\SU(5)$ gauge field
theory as Grand Unified Theory (GUT).
\item[$l\geq6$:] $\SU(l)$
 gauge field theories are developed by physicists as
GUT.
\end{description}
Gelfand, Minlos and Shapiro \cite{gelfand}
considered all relativistic invariant systems of equations of the
form
\[
L^\mu \partial_\mu\psi + i\kappa\psi=0,
\label{eq:gelfand}
\]
where $L^\mu$ are square matrices, $\psi$ -- vector, $\kappa$ --
real constant.

There is an evident generalization of the identity (\ref{1.2})
\[
(i\gamma^\mu \partial_\mu +
m(z\one-y\gamma^5))(i\gamma^\nu \partial_\nu -
m(z\one+y\gamma^5))=-(\dal + m^2),
\label{1.8}
\]
where $z,y$ are complex constants and $z^2+y^2=1$. It leads to the equation
\[
(i\gamma^\mu \partial_\mu-m(z\one+y\gamma^5))\psi=0, \label{1.9}
\]
We can consider the factorizations (\ref{1.2}),(\ref{1.8}) of
the Klein-Gordon operator as
one of possible methods of a reduction of the Klein-Gordon equation of second
order (\ref{1.1})
to a system of equations of first order. There is another method
of such a reduction that leads to the following system of equations of
first order:
\[
i\gamma^\mu \partial_\mu \psi-m( \psi N + \gamma^5 \psi K)=0, \label{1.10}
\]
which depends on two matrices $N,K\in\mlc$. We shall call the
equation (\ref{1.10}) the matrix Dirac equation, emphasizing that
unknown (wave) function $\psi=\psi(x)$ in (\ref{1.10}) is a
$4\!\times\!l$-matrix.
If matrix $\psi$
has only one column, then we get an equation (\ref{1.9}), or standard Dirac
equation (when $N=1, K=0$). The equation (\ref{1.10}), generally
speaking, can't be reduced to the equation of the form
\rr{eq:gelfand}.

\begin{theorem}
If a $4\!\times\!l$-matrix $\psi=\psi(x)$
 with twice continuously
differentiable
elements is a solution of \rr{1.10}, where the matrices
$N,K\in\mlc$
do not depend on $x$ and satisfy the relations
\[
[N,K]=NK-KN=0,\quad N^2+K^2=\one_l, \label{1.11}
\]
where $\one_l$ -- is identity $l\!\times\!l$-matrix,
then the  matrix $\psi$ is also a solution of the Klein-Gordon equation.
\end{theorem}

\proof. Let us consider an action of  the operator
$i\gamma^\mu \partial_\mu$ from left to the
equation \rr{1.10}
$$
(i\gamma^\mu \partial_\mu)^2 \psi - m (i\gamma^\mu \partial_\mu \psi) N +
m\gamma^5(i\gamma^\mu \partial_\mu \psi) K=0,
$$
and use the relations \rr{1.11} and
$$
(i\gamma^\mu \partial_\mu)^2=-\dal, \quad
i\gamma^\mu \partial_\mu \psi=
m(\psi N + \gamma^5\psi K).
$$
As a result we get the Klein-Gordon equation $(\dal+m^2)\psi=0$.
Theorem is proved.
\bigskip

The formula \rr{1.10} gives us a set of equations that depend on two
matrices $N,K$ with the relations \rr{1.11}. How to describe all matrix
pairs $N,K$ that satisfy \rr{1.11}? If we write the matrix $N$ using a Jordan
normal form, then we can find all the  corresponding matrices $K$ by doing
an elementary calculation \cite{gantmacher}.
For example, when $l=4$ we get 15 classes of pairs $N,K$
that depend on several parameters.
\medbreak
\example 1. The matrices
\[
N=V\diag(z_1,\ldots,z_l)V^{-1},\ \
K=V\diag(y_1,\ldots,y_l)V^{-1},\label{1.12}
\]
where ${z_k}^2+{y_k}^2=1$, $V$-- nondegenerate matrix from $\M(l,\C)$,
 satisfy \rr{1.11}.
\medbreak
\example 2. Let $l=4$. The matrices
$$
N=V\pmatrix{z & 1 & 0 & 0\cr
             0 & z & 1 & 0\cr
             0 & 0 & z & 1\cr
             0 & 0 & 0 & z} V^{-1},\quad
K=V\pmatrix{y & a & b & c\cr
             0 & y & a & b\cr
             0 & 0 & y & a\cr
             0 & 0 & 0 & y} V^{-1},
$$
where $z^2+y^2=1,\ y\neq0,\ a=-z/y,\ b=-1/(2y^3),\ c=-z/(2y^5)$
 and $V$-- nondegenerate matrix from $\M(4,\C)$, satisfy \rr{1.11}.
\medbreak
Let us consider a solution $\psi$ of \rr{1.10} with the matrices $N,K$
from the example 1 and connect it with the solutions of the standard Dirac
equation. The columns of the matrix $V$ denote by $v_k\ (k=1,\ldots,l)$.
Then $v_k$ are the
eigenvectors of $N$ corresponding to the  eigenvalues $z_k$.
Simultaneously, $v_k$ are
the  eigenvectors of $K$ corresponding to the  eigenvalues $t_k$. So, if we
multiply \rr{1.10} from right by $v_k$ and denote $\psi_k=\psi v_k$
then we come to $l$ equations
$$
i\gamma^\mu \partial_\mu\psi_k-m(z_k\one+y_k\gamma^5)\psi_k=0,
\quad k=1,\ldots,l
$$
that have the form \rr{1.9}, or \rr{1.6}.

\section{A lagrangian of the matrix Dirac equation.}
The standard Dirac equation \rr{1.6} can be derived from Dirac's
lagrangian
\[
(i/2)(\bar{\psi}\gamma^\mu\partial_\mu\psi
-\partial_\mu\bar{\psi}\gamma^\mu\psi)
-m\bar{\psi}\psi
\label{dirac:lagr}
\]
with the aid of variational principle \cite{bogolubov}.
If $N,K$ are hermitian matrices from $\mlc$, then the matrix
Dirac equation \rr{1.10} can be derived with the aid of
variational principle from the lagrangian
\[
(1/2)\tr\left(i(\bar{\psi}\gamma^\mu\partial_\mu\psi
-\partial_\mu\bar{\psi}\gamma^\mu\psi)
-m(\bar{\psi}\psi N + N\bar{\psi}\psi+ \bar{\psi}\gamma^5\psi K+
K \bar{\psi}\gamma^5\psi)\right),
\label{mdirac:lagr}
\]
in which $\tr$ is a trace of matrix and  $\bar\psi=\psi^\dagger\gamma^0$.

Let us introduce a commutator algebra of matrices $N,K$:
$$
\com(N,K)=\{V\in\M(l,\C) : [V,N]=[V,K]=0\}
$$
and a group
$$
\G(l,N,K)=\com(N,K)\cap\U(l),
$$
where $\U(l)$ is a group of unitary matrices from $\mlc$.
The group $\G(l,N,K)$ is a compact Lie group. A set of all
antihermitian matrices ($V^\dagger=-V$) commuting with $N$ and
$K$
$$
\L(l,N,K)=\com(N,K)\cap\u(l).
$$
can be considered as a real Lie algebra of the Lie group
$\G(l,N,K)$.

The lagrangian \rr{mdirac:lagr} is invariant under
global (not dependent on $x$) transformations
\[
\psi\to\psi V,\quad \bar{\psi}\to V^{-1}\bar{\psi},\ \hbox{with}\ V\in\G(l,N,K),
\label{glob:trans}
\]
This fact is evident from the identity
$\tr(V^{-1}B V)=\tr B$. In order to get a lagrangian which is
invariant under a local (gauge) transformations \rr{glob:trans},
where $V=V(x)$ is a function of $x$ with values in $\G(l,N,K)$,
we must replace in \rr{mdirac:lagr} partial derivatives
$\partial_\mu\psi,\partial_\mu\bar{\psi}$ by the respective
covariant derivatives
$D_\mu\psi=\partial_\mu\psi-\psi a_\mu$,
$\bar{D}_\mu\bar{\psi}=\partial_\mu\bar{\psi}+a_\mu\bar{\psi}$,
that depend on functions
$a_\mu=a_\mu(x)$ with its values in the Lie algebra $\L(l,N,K)$.
A transformation rule for $a_\mu$ has a form
\[
a_\mu\to V^{-1}a_\mu V+V^{-1}\partial_\mu V,\quad V\in\G(l,N,K).
\label{a:trans}
\]

There is a complete gauge invariant lagrangian, which describes a field
$\psi$ interacting with the gauge field $a_\mu$,
\[
\begin{array}{cl}
L=&{1\over2}\tr(i(\bar{\psi}\gamma^\mu(\partial_\mu\psi-\psi a_\mu)
-(\partial_\mu\bar{\psi}+a_\mu\bar{\psi})\gamma^\mu\psi) \\
   &-m(\bar{\psi}\psi N + N\bar{\psi}\psi+ \bar{\psi}\gamma^5\psi K+
K \bar{\psi}\gamma^5\psi))
+{1\over4}\tr(f_{\mu\nu}f^{\mu\nu}),
\end{array}
\label{main:lagr}
\]
where $f_{\mu\nu}=\partial_\mu a_\nu-\partial_\nu a_\mu-[a_\nu,a_\mu]$.
If $N=N^\dagger,\ K=K^\dagger$, then the lagrangian $L$ leads to
the system of equations
\[
\begin{array}{l}
i\gamma^\mu(\partial_\mu\psi-\psi a_\mu)-m(\psi N+\gamma^5\psi K)=0,\\
\partial_\mu a_\nu-\partial_\nu a_\mu-[a_\nu,a_\mu]-f_{\mu\nu}=0,\\
\partial_\mu f^{\mu\nu}-[f^{\mu\nu},a_\mu]=
-\ss{\L}{\pi}(i\bar{\psi}\gamma^\nu\psi),
\end{array}
\label{dym}
\]
$\ss{\L}{\pi} : \u(l)\to\L(l,N,K)$ is a projector operator to the
Lie algebra $\L=\L(l,N,K)$ which can be considered as a linear
subspace of the vector space $\u(l)$ of antihermitian matrices.
The second and the third equations in \rr{dym} are called Yang-Mills
equations.

\medskip
\df. If $V=V(x)$ is a function of $x$ with the value in a Lie
group $\G\subseteq\G(l,N,K)$, then the transformation
\[
\begin{array}{lclcl}
\psi & \to & \psi^\prime & = & \psi V,\\
a_\mu & \to & a_\mu^\prime & = & V^{-1}a_\mu V+V^{-1}\partial_\mu V,\\
f_{\mu\nu} & \to & f_{\mu\nu}^\prime & = & V^{-1}f_{\mu\nu}V
\end{array}
\label{gauge:trans}
\]
is called a gauge transformation of fields
$\psi,a_\mu,f_{\mu\nu}$ with the group $\G$
(note, that the third relation from \rr{gauge:trans} is a consequence
of the second).

\bigskip
The lagrangian \rr{main:lagr} and the system of equations
\rr{dym} are invariant under a gauge transformation
\rr{gauge:trans}. That means
$L(\psi,a_\mu)=L(\psi^\prime,a_\mu^\prime)$, and if $\psi,a_\mu,f_{\mu\nu}$
satisfy \rr{dym}, then
$\psi^\prime,a_\mu^\prime,f_{\mu\nu}^\prime$ from \rr{gauge:trans}
also satisfy \rr{dym}.

\begin{theorem}
Let $N,K\in\mlc$ be such, that
\[
\begin{array}{ccl}
N&=&\alpha_1 \one_l + \beta_1 P_1,\quad P_1 P_1^\dagger=\one_l,\quad P_1^\dagger=P_1,
\quad \alpha_1,\beta_1\in\R\\
K&=&\alpha_2 \one_l + \beta_2 P_2,\quad P_2 P_2^\dagger=\one_l,\quad P_2^\dagger=P_2,
\quad \alpha_2,\beta_2\in\R,
\end{array}
\label{conserv:cond}
\]
and $\psi=\psi(x)$ is a solution of the matrix Dirac equation
\rr{1.10}. Let us denote
 $J^\nu=\ss{\L}{\pi}(i\bar{\psi}\gamma^\nu\psi)$, where
$\L=\L(l,N,K)=\com(N,K)\cap\u(l)$. Then
\[
\partial_\nu J^\nu(\psi)=0.
\label{current}
\]
\end{theorem}

In other words, if we pose additional conditions
\rr{conserv:cond} on $N,K$, then a right hand part of Yang-Mills
equations is a current of matrix Dirac equation.
\bigskip

\proof. A solution $\psi$ of the matrix Dirac equation \rr{1.10}
also satisfies an identity
\[
\partial_\mu(i\bar\psi \gamma^\mu \psi)-
m(\bar\psi\psi N-N^\dagger \bar\psi\psi+\bar\psi \gamma^5\psi K-
K^\dagger \bar\psi \gamma^5\psi)=0.\label{int:ident}
\]
which is consequence of \rr{1.10}. To show this we must multiply
\rr{1.10} from left on $\bar\psi$ and subtract hermitian
conjugated equation
$-i\partial_\mu\psi^\dagger  (\gamma^\mu)^\dagger - m(N^\dagger \psi^\dagger -
K^\dagger \psi^\dagger \gamma^5)=0$,
multiplied from right on $\gamma^0\psi$. The result can be
written in the form \rr{int:ident}.

Let us prove that if $N,K$ satisfy \rr{conserv:cond}, then
\[
\ss{\L}\pi(\bar\psi\psi N-N^\dagger \bar\psi\psi+\bar\psi \gamma^5\psi K-
K^\dagger \bar\psi \gamma^5\psi)=0.\label{3.15}
\]
It is easy to check, that the matrices
$$
B_1=\bar\psi\psi N-N^\dagger \bar\psi\psi,\quad
B_2=\bar\psi \gamma^5\psi K-K^\dagger \bar\psi \gamma^5\psi.
$$
anticommute with the matrices $P_1,P_2$ respectively:
$$
B_1 P_1=-P_1 B_1,\quad B_2 P_2=-P_2 B_2.
$$
We can rewrite this fact as
\[
B_1\in\mlc \setminus  \com(P_1),\quad
B_2\in\mlc \setminus  \com(P_2),\label{3.16}
\]
where $\com(P)\subseteq\mlc$ is a subspace of all matrices
commuting with the matrix $P\in\mlc$. It follows from \rr{3.16},
that
\[
B_1+B_2\in\mlc\setminus (\com(P_1)\cap\com(P_2)).
\label{3.17}
\]
Taking into account the definition of the Lie algebra $\L$, we
get from \rr{3.17} the relation
$\ss{\L}\pi(B_1+B_2)=0$.
Acting by the projector operator  $\ss{\L}{\pi}$ on the right
hand and left hand parts of the identity \rr{int:ident}, and
using the formula \rr{3.15}, we get
$$
\ss{\L}{\pi}(\partial_\mu(i\bar\psi \gamma^\mu\psi))=
\partial_\mu J^\mu=0.
$$
Theorem is proved.
\bigskip

\section{A general form of the matrices $N,K$.}
The matrix Dirac equation \rr{1.10} depends on two matrices $N,K$.
Let us compile all conditions on the matrices $N,K$.

If $N,K$ satisfy \rr{1.11}, then the solution $\psi$ of equation
\rr{1.10}  also satisfies Klein-Gordon equation.

If $N,K$ are hermitian, then the matrix Dirac equation can be
derived from the lagrangian \rr{mdirac:lagr} with the aid of
variational principle.

If $N,K$ satisfy \rr{conserv:cond},  then the right hand part of
Yang-Mills equations \rr{dym} is a current of matrix Dirac
equation.

Evidently, if $N,K$ satisfy \rr{conserv:cond}, then $N,K$ are
hermitian matrices.

\begin{theorem}
The matrices $N,K\in\mlc$ satisfy \rr{1.11} and \rr{conserv:cond}
if and only if
\[
\begin{array}{ccl}
N & = & U^\dagger\diag(\underbrace{\cos\xi,\ldots,\cos\xi}_{\hbox{$p$ pieces}},
\underbrace{\cos\eta,\ldots,\cos\eta}_{\hbox{$q$ pieces}})U,\\
K & = & U^\dagger\diag(\underbrace{\sin\xi,\ldots,\sin\xi}_{\hbox{$p$ pieces}},
\underbrace{\sin\eta,\ldots,\sin\eta}_{\hbox{$q$ pieces}})U,
\end{array}
\label{NK:canon1}
\]
or
\[
\begin{array}{ccl}
N & = & U^\dagger\diag(\pm1,\ldots,\pm1)U\cos\xi,\\
K & = & U^\dagger\diag(\pm1,\ldots,\pm1)U\sin\xi,
\end{array}
\label{NK:canon2}
\]
where $p+q=l$, $U$-unitary matrix and  $0\leq\xi,\eta<2\pi$.
\end{theorem}

\proof. The matrix $N$ in \rr{conserv:cond} depends on a hermitian
and simultaneously unitary matrix $P_1$. So, $P_1$ can be written
with the aid of a unitary matrix $U$:
$$
P_1=U^\dagger\diag(\pm1,\ldots,\pm1)U.
$$
Hence
$$
N=U^\dagger\diag(\lambda_1,\ldots,\lambda_l)U,
$$
where $\lambda_k=\lambda^+=\alpha_1+\beta_1$, or
$\lambda_k=\lambda^-=\alpha_1-\beta_1$.
The same is true for the matrix $K$:
$$
K=V^\dagger\diag(\epsilon_1,\ldots,\epsilon_l)V,
$$
where $V^\dagger V=\one$,
$\epsilon_k=\epsilon^+=\alpha_2+\beta_2$, or
$\epsilon_k=\epsilon^-=\alpha_2-\beta_2$. Matrices $N,K$
commute with each other and can be simultaneously reduced to a
diagonal form with the same similarity transformation
\cite{gantmacher}. That means, that we can take $U=V$.
The condition $N^2+K^2=\one$ gives
${\lambda_k}^2+{\epsilon_k}^2=1,\ k=1,\ldots,l$. There are two
possibilities. The first: $(\lambda^+)^2\neq(\lambda^-)^2$,
$(\epsilon^+)^2\neq(\epsilon^-)^2$ and so
$$
\lambda_k=\lambda^+=\cos\xi,\ \epsilon_k=\epsilon^+=\sin\xi,\
k=1,\ldots,p,
$$
$$
\lambda_k=\lambda^-=\cos\eta,\ \epsilon_k=\epsilon^-=\sin\eta,\
k=p+1,\ldots,l.
$$
The second: $(\lambda^+)^2=(\lambda^-)^2$,
$(\epsilon^+)^2=(\epsilon^-)^2$ and so
$$
\lambda_k=\pm\cos\xi,\ \epsilon_k=\pm\sin\xi,\ k=1,\ldots,l.
$$
Theorem is proved.
\bigskip

\section{A polar gauge.}
There is a theorem about the polar decomposition of matrix.
\begin{theorem}[\cite{horn}]
An arbitrary matrix $M\in\mlc$ can be written in a form $M=PU$,
where $P\in\mlc$ is a hermitian nonnegative defined matrix with
the same rank as $M$, and a matrix $U\in\mlc$ is unitary.
\end{theorem}

Let us consider a system of equations \rr{dym} where
 $l=4,\ N=\one\cos\xi,\
K=\one\sin\xi$ with the gauge group $\G=\U(4)$. And let
$\psi,a_\mu,f_{\mu\nu}$ be a continuously differentiable
functions of $x$ which satisfy \rr{dym} and such that
$\psi\in\M(4,\C)$ and $a_\mu,f_{\mu\nu}\in\L=\u(4)$.
Using the theorem 4, we get that in every point $x\in\R^4$
the matrix $\psi=\psi(x)$ can be written in a form $\psi=PU$,
where $P\in\mlc$ is a hermitian nonnegative defined matrix and
$U\in\mlc$ is a unitary matrix.

Let us suppose, that the matrix $\psi=\psi(x)$ can be written in
a form $\psi(x)=P(x) U(x)$ in some region
$\Omega\subset\R^4$ and $P(x), U(x)$ have continuously
differentiable elements for all $x\in\Omega$. In that case the
solution
$\psi=\psi(x)$ of \rr{dym} in the region $\Omega$ defines a
unitary matrix $U=U(x)$ with continuously differentiable
elements.
We can take $V=V(x)=U^{-1}$. The system of equations \rr{dym} is
invariant under a gauge group $\U(4)$, and after a gauge
transformation with $V\in\U(4)$, we come to the solution of
\rr{dym}
$$
\psi^\prime=\psi V,\quad
a_\mu^\prime=V^{-1}a_\mu V+V^{-1}\partial_\mu V,\quad
F^\prime=V^{-1} F V,
$$
where $\psi^\prime=\psi^\prime(x)$ is a hermitian nonnegative
defined matrix and we name it solution of \rr{dym} in polar gauge.

In quantum mechanics particles are described by the wave
functions which belong to some complex finite or infinite
dimensional vector (hilbert) space. Observables are hermitian
operators on that space. It will be naturally to suppose, that
the solution $\psi$ of \rr{dym} is a wave function of fermion.

If we take $\psi$ in the polar gauge (in the case of gauge group
$\U(4)$), then $\psi$ is a hermitian \4-matrix equivalent to the
hermitian $16\!\times\!16$-matrix $\Psi\otimes{\bf1}$. That
means, that we can consider $\psi$ not only as a wave function, but
simultaneously as some observable value.

\vskip 5mm
Let us make two final remarks.

It can be shown, that the matrix Dirac equation \rr{1.10} is
invariant under the Lorenz transformations.

In case $l=4$, the wave function $\psi$, which satisfy \rr{1.10},
is a \4-matrix and so, it can be represented as a linear
combination of basis vectors \rr{1.5} of Dirac's algebra
(Clifford's algebra). Substituting it into \rr{1.10}, we get an
equation written in Clifford's algebra terms and not dependent on
matrix representations. Such an equation can be naturally
generalized to the curved space-time with an arbitrary
pseudorimanian metric.

This questions will be considered in the next publications.

\vskip 5mm
We thank Prof. A.A.Dezin, Prof. V.V.Zarinov and Prof. A.A.Slavnov
for fruitful discussions.



\begin{thebibliography}{99}
\bibitem{dirac} Dirac P.A.M. Proc. Roy. Soc. Lond. A117 (1928) 610.
\bibitem{weil}  Weil H. ZS. f.Phys. 56 (1929) 330.
\bibitem{fock}  Fock V. Journ. de Physique 10 (1929) 392.
\bibitem{yang} Yang C.N., Mills R.L. Phys. Rev. 96 (1954) 191.
\bibitem{marchuk1}  Marchuk N.G. Differential Equations (in Russian),
4 (1984) 653.
\bibitem{marchuk2} Marchuk N.G. Preprint CTS of IISc, 1 (1994), Bangalore.
\bibitem{hestenes} Hestenes D. Space-Time Algebra, Gordon and Breach, New York 1966.
\bibitem{kahler} K\"ahler E. Randiconti di Mat. (Roma) ser. 5, 21, (1962) 425.
\bibitem{pestov} Pestov A.B. Preprint P2-5798, Dubna 1971.
\bibitem{pezzaglia} Pezzaglia W.M.jr., Differ A.W. Proc. of XXIIth
    International Conf. on
    "Differential Geometric Methods in Theoretical Physics",
    Ixtapa-Zihuatanejo, Mexico, 1993, p. 437-446.
\bibitem{georgi} Georgi H., Glashow S.L. Phys. Rev. D6 (1972) 429.
\bibitem{gelfand} Gelfand I.M., Minlos R.A. and Shapiro Z.Ya.
        Representations of the Rotations and Lorentz Groups
        and their Applications,
        Pergamon, New York, 1963.
\bibitem{gantmacher} Gantmacher F.R. The Theory of Matrices,
Chelsea Publishing, New York, 1959.
\bibitem{bogolubov}Bogoliubov N.N. and Shirkov D.V. Introduction
to the Theory of Quantized Fields. Interscience, New York and
London, 1959.
\bibitem{horn} Horn R.A., Johnson C.R. Matrix Analysis.
        Cambridge Univ. Press 1986.
\bibitem{marchuk3} Marchuk N.G. Advances in Applied Clifford Algebras, 8,
No.1,(1998),181-224.
\end{thebibliography}
\end{document}